\documentclass{article}
\usepackage{spconf,amsmath,graphicx,hyperref}
\usepackage{amssymb}
\usepackage{cases}
\usepackage{bm}

\title{BeamFocusNet: Beamforming-based Explicit Spatial Signal Focusing for Robust DoA Estimation under Low-SNR and Single-Snapshot Conditions}
\name{$\text{Xuyao Deng}$, $\text{Qisheng Xu}$, $\text{Shuo Liu}$, $\text{Yong Dou}$, $\text{Wen Zhang}$, $\text{Kele Xu}^*$\thanks{*Corresponding author. kele.xu@ieee.org}}
\address{National University of Defense Technology, Changsha, China}
\begin{document}
%
\maketitle
\begin{abstract}
DoA estimation plays a crucial role in signal processing. Inspired by beamforming, recent works employ neural networks to estimate filters for filtering received signals to achieve DoA estimation. These methods typically generate filters by implicitly focusing signals and suppressing noise. However, this couples the two objectives, making training difficult to balance and resulting in poor robustness, especially under low SNR and limited snapshots. To address this issue, we propose the BeamFocusNet method, which generates filters through explicit signal focusing, introducing a new paradigm for neural network-based filter generation. Extensive experiments under various challenging conditions demonstrate the superiority and robustness of the proposed method in DoA estimation. Code is available at \url{https://github.com/colaudiolab/BeamFocusNet}.
\end{abstract}

\vspace{-0.2cm}
\section{Introduction}
\label{sec:intro}
\vspace{-0.25cm}
Direction-of-Arrival (DoA) estimation is a core task in array signal processing~\cite{benesty2017fundamentals}. Traditional model-driven methods such as MUSIC~\cite{schmidt1986multiple} and MVDR~\cite{capon2005high} perform well under ideal conditions but degrade significantly under coherent signals, limited snapshots, or low SNR. Recent deep learning approaches—categorized into pure data-driven~\cite{papageorgiou2021deep,gao2023gridless,zhu2019deep,chen2020deep,cong2020robust,de2022resnet,chen2024sdoa,11462975,11462557}, model-driven inspired~\cite{merkofer2023music,merkofer2022deep,shmuel2023deep,deng2025beamformnet,ji2024transmusic}, and neural-enhanced~\cite{shmuel2024subspacenet,lee2022deep,gast2025dcd,11463385}—have improved robustness in these challenging scenarios. Among them, BeamformNet achieves interpretable DoA estimation by approximating the optimal spatial filter through implicit signal focusing and noise suppression. Despite the significant achievements of BeamformNet, its framework still suffers from three fundamental limitations that hinder performance improvement under extreme conditions.

 First, BeamformNet implicitly learns spatial filters by coupling two subtasks—signal focusing and noise suppression—which forces a trade-off between enhancing the signal and suppressing noise, leading to poor performance in low SNR environments. Second, since the filter operates directly on the array received signals, it becomes extremely difficult to infer spatial information from observations when the number of snapshots is limited. Third, BeamformNet directly takes the array received signal $\bm{X}$ as input, causing the dimensionality of its feature space and network architecture to be tied to the number of snapshots. Under a limited number of snapshots, this results in severe dimensional insufficiency and degraded spatial resolution.

To address the above challenges, this paper proposes a novel model-driven deep learning framework named BeamFocusNet. Through theoretical analysis, we anchor the optimization objective to the sparse structure of the product of the filter $\bm{B}$ and the steering vector dictionary $\bm{A}$, and by leveraging neural networks to learn the beamforming process and generate new spatial filters, we achieve a more robust DoA estimation under extreme conditions. Our core improvements lie in input reconstruction and explicit spatial signal focusing via target decoupling. We make three contributions: First, we propose an innovative explicit supervision paradigm that focuses the signal via a sparse focusing structure, avoiding explicit noise modeling and direct noise contamination of the supervision target. Second, we utilize covariance matrix to resolve input and network dimensionality collapse under single snapshot and enabling inference independent of snapshots. Third, extensive challenging experiments show that BeamFocusNet exhibits robustness and superiority, particularly in single-snapshot and low-SNR scenarios.

\begin{figure}[htb]
\begin{minipage}[b]{0.44\linewidth}
  \centering
  \centerline{\includegraphics[width=4.4cm]{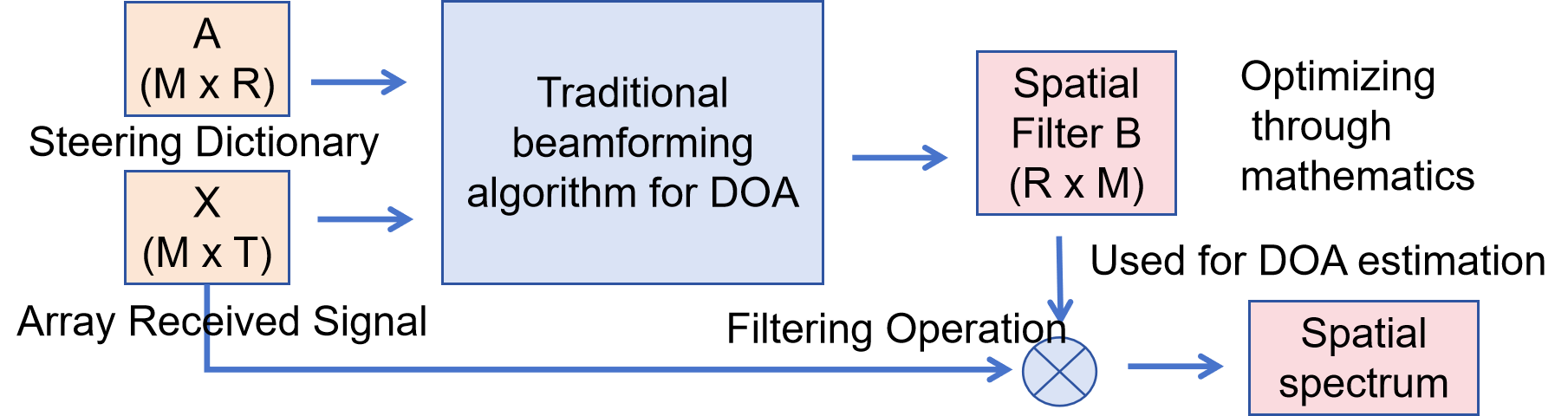}}
  \centerline{(a)}\medskip
\end{minipage}
\begin{minipage}[b]{0.46\linewidth}
  \centering
  \centerline{\includegraphics[width=4.4cm]{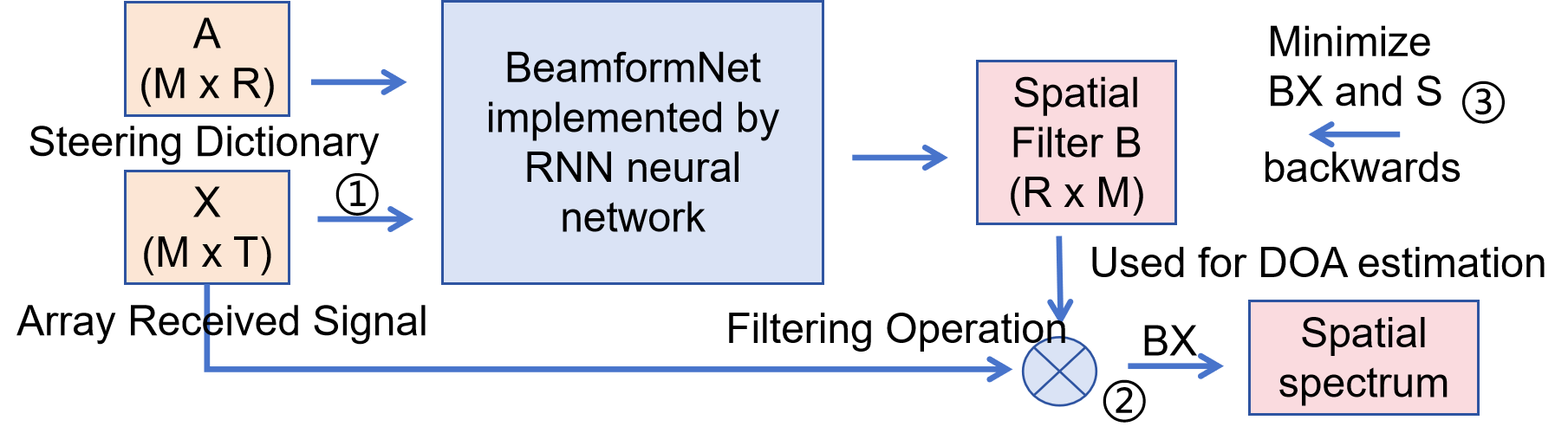}}
  \centerline{(b)}\medskip
\end{minipage}
\begin{minipage}[b]{0.44\linewidth}
  \centering
  \centerline{\includegraphics[width=4.4cm]{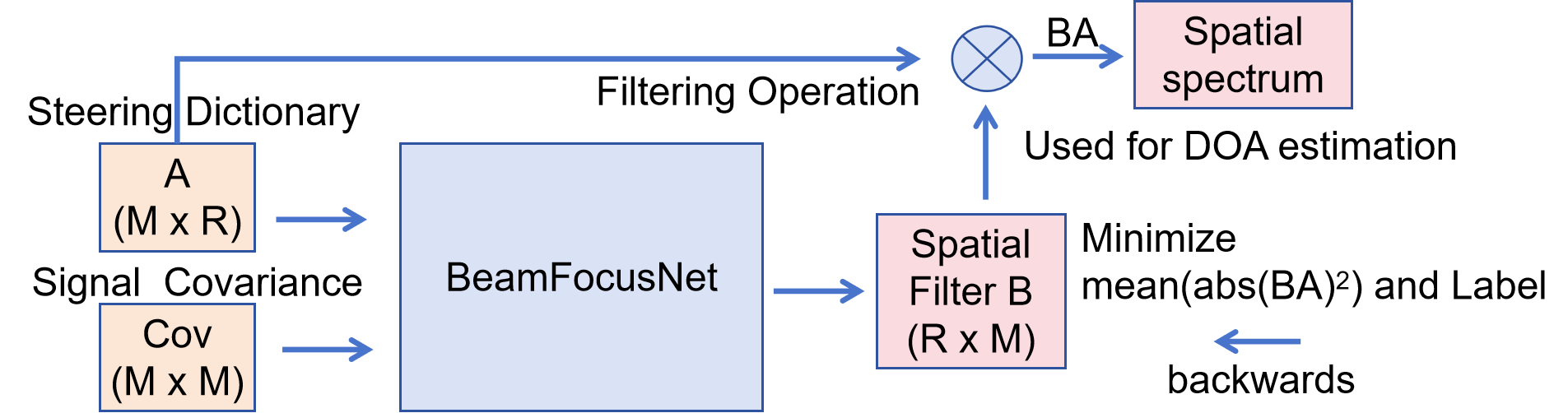}}
  \centerline{(c)}\medskip
\end{minipage}
\hfill
\begin{minipage}[b]{0.46\linewidth}
  \centering
  \centerline{\includegraphics[width=3.9cm]{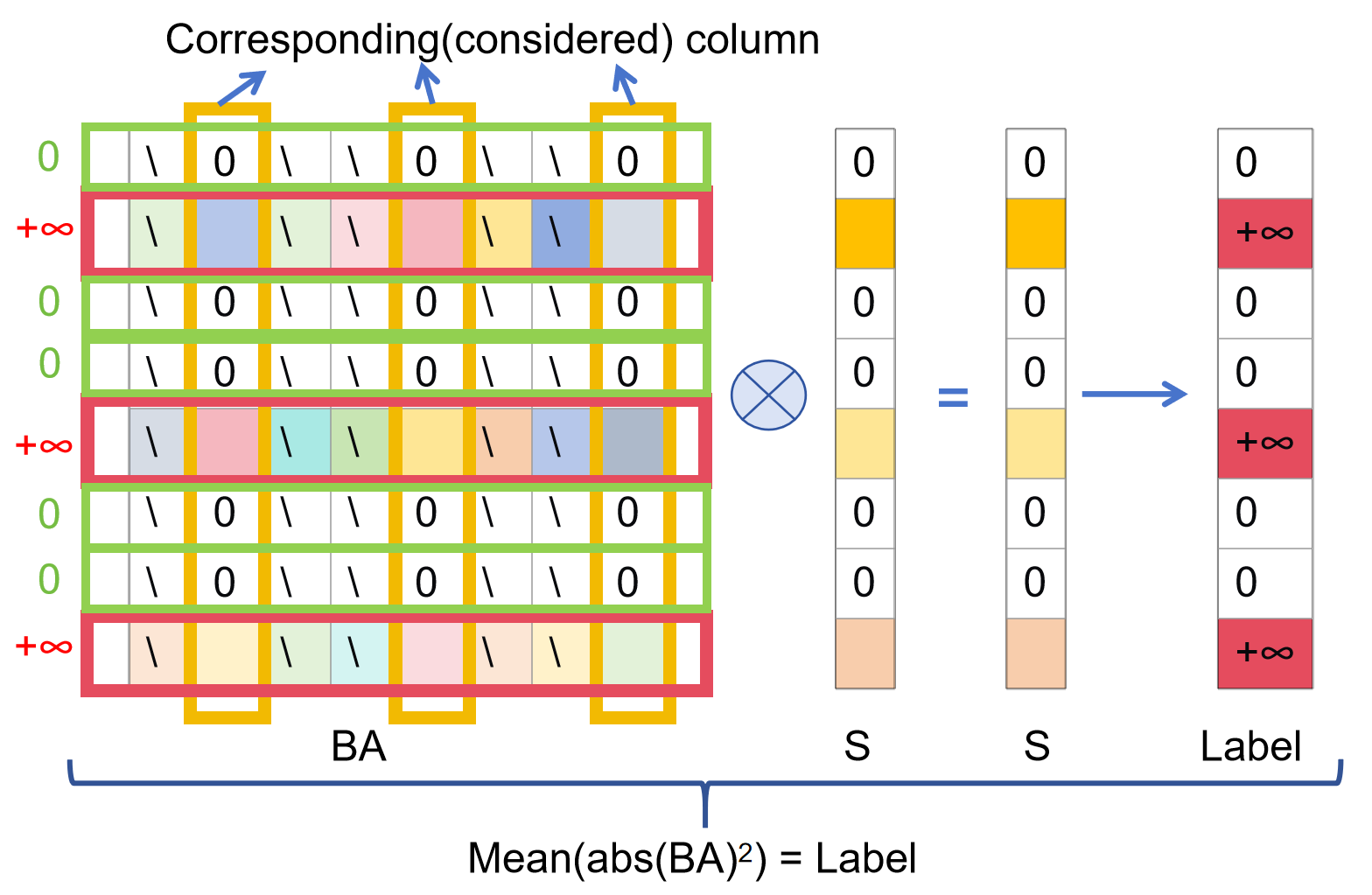}}
  \centerline{(d)}\medskip
\end{minipage}
\vspace{-0.5cm}
\caption{Subfigures (a)–(c) show DoA estimation flowcharts for conventional beamforming, BeamformNet, and BeamFocusNet. Subfigure (d) illustrates the special structure of $\bm{BA \times S}=\bm{S}$.}
\label{fig:process}
\end{figure}

\section{Sparse Signal model and motivations}
\label{sec:signal_model}
\vspace{-0.25cm}
\subsection{Sparse Signal model}
\vspace{-0.15cm}
Consider $K$ narrowband far-field signals impinging on a $M$-element array. Under ideal conditions, the received spatially sparse signal at time $t$ is modeled as:
\vspace{-0.2cm}
{\small \begin{equation}
\label{eq:x=as+n_complete}
\begin{bmatrix}
x_1(t) \\
x_2(t) \\
\vdots \\
x_M(t)
\end{bmatrix}=
\begin{bmatrix}
\bm{a}_{1}(\omega_0)...\bm{a}_{R}(\omega_0)
\end{bmatrix}
\begin{bmatrix}
s_1(t) \\
s_2(t) \\
\vdots \\
s_R(t)
\end{bmatrix}+
\begin{bmatrix}
n_1(t) \\
 \\
n_2(t) \\
\vdots \\
n_M(t)
\end{bmatrix}
\vspace{-0.2cm}
\end{equation}}of which
{\small \begin{equation}
    \label{eq:s_i(t)=0_s_i(t)=s_i(t)}
    \begin{cases}s_i(t)=s_i(t), & i \in Grids~with~signals~present \\s_i(t)=0, & i \in other~grids
    \end{cases}
\end{equation}}
{\small
\begin{equation}
\label{eq:bf(a)_i(w_0)}
\bm{a}_i(\omega_0)=
\begin{bmatrix}
e^{-j\omega_0\tau_{1i}} \\
e^{-j\omega_0\tau_{2i}} \\
\vdots \\
e^{-j\omega_0\tau_{Mi}}
\end{bmatrix}\quad i=1,2,...,K
\end{equation}}
and
{\small
\begin{equation}
    \label{eq:w0=2pif}
    \omega_0 = 2\pi f = \frac{2\pi c}{\lambda},
\end{equation}}where $s_i(t)$ is the $i$-th source signal, $x_j(t)$ is the signal received by the $j$-th sensor, $n_j(t)$ is the additive noise at the $j$-th sensor, $\tau_{ji}$ is the time delay of the $i$-th signal at the $j$-th sensor relative to the reference, $R$ is the number of grids at resolution $\delta$, and $f$, $c$, and $\lambda$ denote frequency, wave speed, and wavelength, respectively.

Following prior work, we consider a half-wavelength uniform linear array (ULA), thus $\tau_{ji}$ is given by:
\vspace{-0.2cm}
{\small\begin{equation}
    \label{eq:tau_ji_in_1D}
    \tau_{ji}= \frac{1}{c} \left( y_j \sin \theta_i  \right).
\end{equation}}where $y_j$ are the y-coordinates of the $j$-th sensor with the reference element at the origin, and $\theta_i$ is the azimuth angle of the $i$-th signal. we refer the reader to references~\cite{Wang2004Spatial,tuncer2009narrowband,friedlander2009wireless,chen2010introduction,deng2025beamformnet} for further details. The discrete sparse model in \eqref{eq:x=as+n_complete} can then be written in matrix form as:
\vspace{-0.2cm}
{\small \begin{equation}
\label{eq:X=A_complete S+N}
\bm{X}=\bm{AS}+\bm{N}.
\vspace{-0.2cm}
\end{equation}}

DoA estimation is the inverse of signal reception: given $T$ snapshots from an $M$-element ULA, it aims to accurately recover the azimuth angles $\theta_i$ of $K$ sources, assuming $K < M$ and typically $K$ are unknown.

\subsection{Motivations}
\vspace{-0.15cm}
The flowchart of traditional beamforming for DoA estimation is shown in Fig.~\ref{fig:process}(a): it defines an optimization term and derives filter $\bm{B}$ mathematically, which is then applied to the received signal $\bm{X}$ to generate a spatial spectrum. BeamformNet (Fig.~\ref{fig:process}(b)), on the other hand, learns $\bm{B}$ by minimizing the loss between $\bm{BX}$ and $\bm{S}$ via a neural network that implicitly learns spatial signal focusing ($\bm{BAS}=\bm{S}$) and noise suppression ($\bm{BN}=\bm{0}$). Their procedures can be summarized as follows: different beamforming techniques can be viewed as deriving distinct spatial filters $\bm{B}$ based on different objectives and then applying them to the array's received signals $\bm{X}$ to estimate the DoA~\cite{van1988beamforming}, that is:
\vspace{-0.2cm}
{\small \begin{equation}
    \label{eq:BX=BAS+BN}
    \bm{BX} = \bm{BAS} + \bm{BN}.
\vspace{-0.2cm}
\end{equation}}

However, this framework has three fundamental limitations that restrict its performance under extreme conditions. First (corresponding to \textcircled{1} in Fig.~\ref{fig:process}(b)), regarding input representation and network adaptability, BeamformNet takes the snapshot-dependent signal matrix as input. In single-snapshot cases, the temporal dimension collapses, so the first RNN layer receives only a 1D vector, losing temporal context and limiting spatial resolution. Moreover, the linear layers in it also depend on snapshot count; with one snapshot, the input feature dimension is insufficient, further degrading resolution. Second (corresponding to \textcircled{2}), it applies filters $\bm{B}$ directly in the element-snapshot domain within $\bm{X}$, where signal and noise are highly intermingled. This makes it extremely difficult to retrieve spatial information from a limited number of observations (e.g., a single snapshot) because the embedded phase and amplitude information are easily contaminated by noise and suffer from inherent ambiguity. Third (corresponding to \textcircled{3}), in terms of learning objective and signal representation, it's objective implicitly couples two sub-tasks—signal focusing and noise suppression. In low-SNR environments, noise significantly disturbs the optimization direction and estimation accuracy of the filters, forcing a trade-off between signal enhancement and noise suppression.

To address the above challenges, we propose a novel learning paradigm named BeamFocusNet, whose flowchart is shown in Fig.~\ref{fig:process}(c). Unlike BeamformNet, we shift the objective of the learned filter from $\bm{BX}=\bm{S}$ to a special structure on $\bm{BA}$. This is because, under the sparse signal model, the true source vector $\bm{S}$ is inherently sparse with unknown numerical values but known source direction labels, causing the product $\bm{BA}$ of the ideal spatial filter $\bm{B}$ and the steering dictionary $\bm{A}$ to exhibit a special structure (as illustrated in Fig.~\ref{fig:process}(d)) in order to meet $\bm{BAS}=\bm{S}$. Its corresponding columns (corresponding to the yellow box in Fig.~\ref{fig:process}(d)) yield significant responses at the angle grids corresponding to actual sources, while approaching zero elsewhere. Other column structures are not restricted due to the corresponding 0 value in $\bm{S}$. This structure is directly defined in the spatial domain, fully encodes spatial focusing information, and is independent of noise. Therefore, shifting the learning objective from the filtered output to this sparse spatial structure can fundamentally avoid explicit noise modeling, enabling pure “spatial signal focusing”. Moreover, the steering dictionary $\mathbf{A}$ is unaffected by snapshots. These stabilize training under low SNR and bypasses spatial inference from limited snapshot observations within $\bm{X}$. Now, We analyze the existence of a filter $\bm{B}$ that yields the structure of $\bm{BA}$ satisfying $\bm{BAS}=\bm{S}$. Since $\bm{S}$ is sparse, we only consider positions where sources exist (the three yellow blocks in $\bm{S}$ in Fig.~\ref{fig:process}(d)) . These positions multiply and sum with corresponding columns in each row of $\bm{BA}$; other columns are ignored because $\bm{S}$ contains zeros there. As the steering dictionary $\bm{A}$ is known, this imposes $K$ constraints on each row of $\bm{B}$, whose degree of freedom is $M$. Under the narrowband assumption and $K<M$, such row vectors always exist, so $\bm{B}$ forming the structure of $\bm{BA}$ exists. Based on this, we depart from the conventional beamforming mindset of applying $\bm{B}$ directly to $\bm{X}$. Instead, we qualitatively justify applying $\bm{B}$ to $\bm{A}$ and anchor the optimization objective to the sparse structure of $\bm{BA}$: We aim to maximize the values of the rows in $\bm{BA}$ (corresponding to the red box and $+\infty$ in Fig.~\ref{fig:process}(d)) that correspond to signals present in $\bm{S}$ and set the other rows to zero (corresponding to the green box in Fig.~\ref{fig:process}(d)). For the convenience of training, it corresponds to the average row energy in Eqs.(\ref{eq:P'=BA})-(\ref{eq:P=Mean(Abs)}), and to the training objective that drives $tanh(\bm{P})$ toward 1 in signal regions and toward 0 in other regions. Notably, BeamFocusNet generates data-adaptive filters $\bm{B}$ to accommodate the distinct $\bm{BA}$ structures corresponding to different $\bm{S}$ in $\bm{BAS}=\bm{S}$.

To address the problem of insufficient input feature dimension and snapshot-dependent layer scaling under limited snapshots, we replace inputs $\bm{X}$ with the sample covariance matrix $\bm{Cov}=\bm{XX}^H \in \mathbb{C}^{M \times M}$. This substitution fixes the network input dimension to $M \times M$, fully decoupling it from the snapshot count $T$, fundamentally eliminating the risk of input and layer dimension collapse in the single-snapshot scenario. In addition, it provides the network with second-order statistical information about the array manifold. 

\begin{figure*}[t] 
\vspace{-0.4cm}
	\centering
	\includegraphics[width=7in]{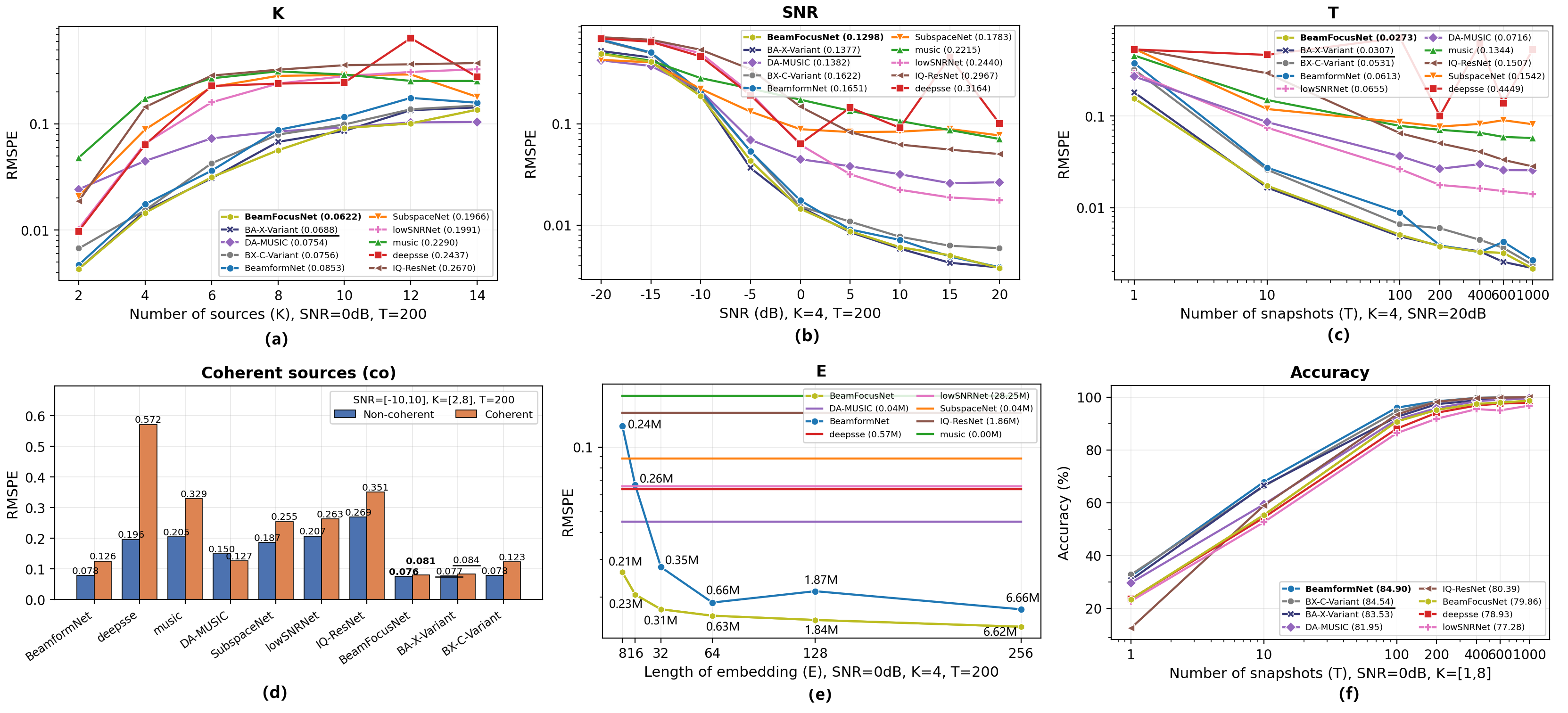}
    \vspace{-0.9cm}
	\caption{Experimental results. The best and second-best performances are denoted by bold and underline. Numbers in parentheses denote the average performance. BeamFocusNet demonstrates more robust and superior performance in DoA estimation.}
	\label{fig:Experiments}
\vspace{-0.3cm}
\end{figure*}

\begin{figure}[t] 
	\centering
	\includegraphics[width=3.3in]{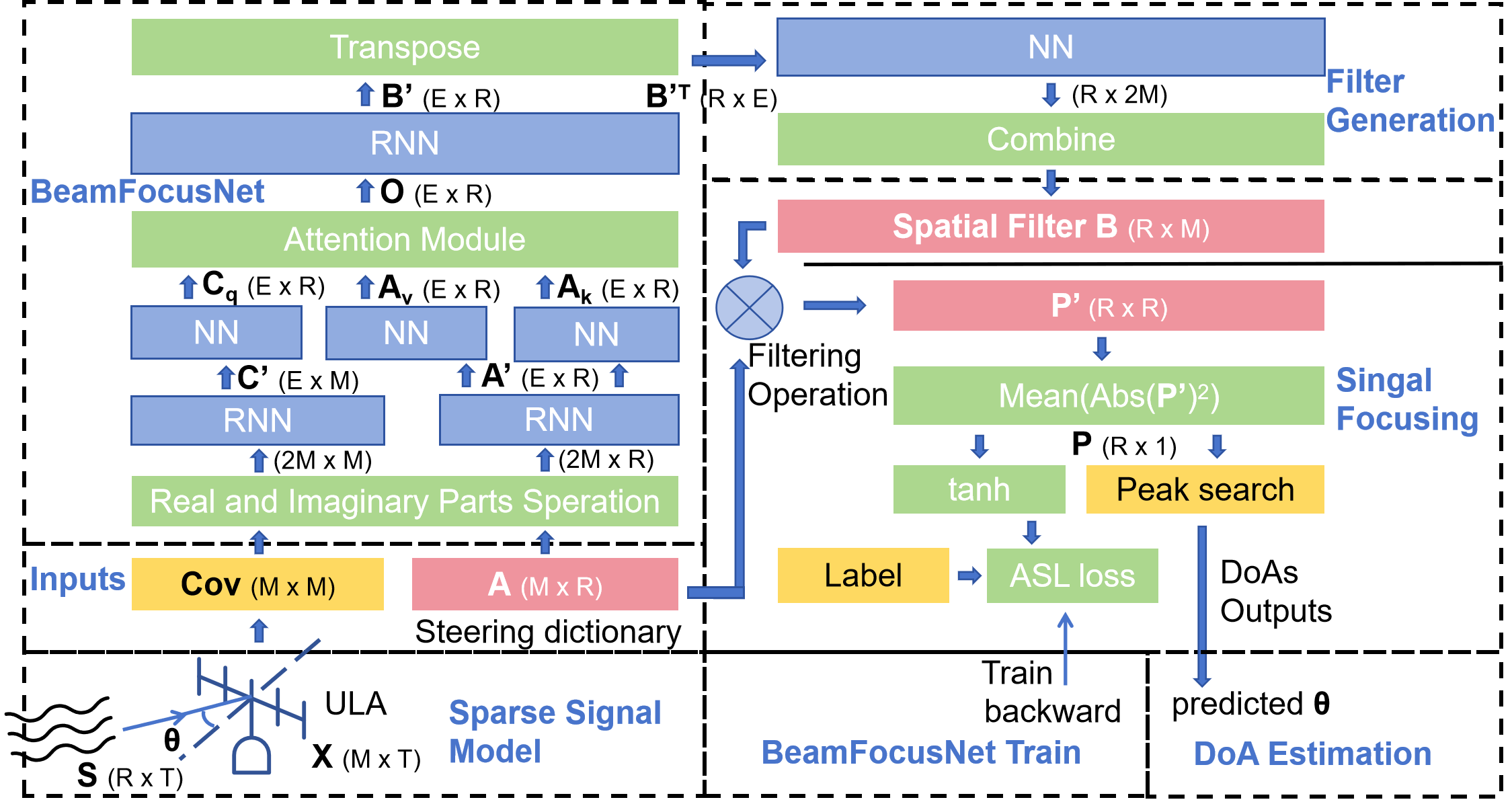}
    \vspace{-0.35cm}
	\caption{The overall structure of BeamFocusNet.}
	\label{fig:BA-BeamformNet}
\vspace{-0.5cm}
\end{figure}

\vspace{-0.3cm}
\section{BeamFocusNet}
\label{sec:BA-BeamformNet}
\vspace{-0.25cm}

To ensure a fair comparison with BeamformNet, the backbone of BeamFocusNet follows that of BeamformNet, as shown in Fig.~\ref{fig:BA-BeamformNet}. The network takes the steering dictionary $\bm{A}$ and signal covariance $\bm{Cov}$ as inputs, separates their real and imaginary parts, concatenates, and feeds them into two different bidirectional RNNs to obtain output with dimension~$E$:
\vspace{-0.1cm}
\begin{numcases}{}
    \bm{A'} = RNN(concat(separate(\bm{A}))) \\
    \bm{C'} = RNN(concat(separate(\bm{Cov})))
\vspace{-0.2cm}
\end{numcases}with 
{\small \begin{equation}
\begin{gathered}
    \Im(\cdot),~\Re(\cdot) = separate(\cdot)\\
    [\Im(\cdot),~\Re(\cdot)]^T = concat(\Im(\cdot),\Re(\cdot)),
\end{gathered}
\end{equation}}where $\Im(\cdot)$ and $\Re(\cdot)$ denote the imaginary and real parts of each matrix element, respectively, to form a new matrix.

Then the network employs three distinct linear networks to transform $\bm{A'}$ and $\bm{C'}$ into vectors $\bm{A}_v$, $\bm{A}_k$, and $\bm{C}_q$ of the same dimension, serving as the value, key, and query in the subsequent attention mechanism. The process is:
\vspace{-0.2cm}
{\small \begin{equation}
\begin{gathered}
    \bm{A}_v = NN(\bm{A'})\\
    \bm{A}_k = NN(\bm{A'})\\
    \bm{C}_q = NN(\bm{C'}).
\end{gathered}
\end{equation}}
\vspace{-0.2cm}
{\small \begin{equation}
    \bm{O} = softmax(\frac{\bm{C}_q{\bm{A}_k}^T}{\sqrt{R}})\bm{A}_v.
\end{equation}}

The attention mechanism retrieves target spatial information from the steering dictionary using the input data, producing $\bm{O}$ as the precursor of the input-adaptive filter $\bm{B}$. An RNN and a linear network then extract the real and imaginary parts of $\bm{B}$, which are combined to form the final filter $\bm{B}$:
\vspace{-0.2cm}
{\small \begin{equation}
    \bm{B'} = RNN(\bm{O}).
\end{equation}}
\vspace{-0.5cm}
{\small \begin{equation}
\begin{gathered}
    [\Im(\bm{B}),\Re(\bm{B})]^T = NN(\bm{B'}^T)\\
    \bm{B} = combine([\Im(\bm{B}),\Re(\bm{B})]^T) = \Re(\bm{B}) + j \cdot \Im(\bm{B}).
\end{gathered}
\end{equation}}

We then apply the filter $\bm{B}$ to the steering dictionary $\bm{A}$ to obtain $\bm{BA}$. To enforce the unique structure in Fig.~\ref{fig:process}(d) such that $\bm{BAS}=\bm{S}$, we take the column-wise average of its energy. The specific process is:
\vspace{-0.2cm}
{\small \begin{equation}
    \bm{P'} = \bm{BA}.
    \label{eq:P'=BA}
\end{equation}}
\vspace{-0.6cm}
{\small \begin{equation}
    \bm{P} = Mean(Abs(\bm{P'})^2).
    \label{eq:P=Mean(Abs)}
\end{equation}}$Abs$ denotes element-wise modulus, and $Mean$ denotes averaging over the column dimension. We then obtain the energy spectrum $\bm{P}$, where the goal is to maximize energy at source directions and minimize it elsewhere. For training, we normalize $\bm{P}$ to [0,1] using the tanh activation and employ binary labels \{0,1\}. To mitigate class imbalance from sparse signals, we adopt the Asymmetric Loss (ASL)~\cite{ridnik2021asymmetric} as the objective function. 

To filter weak spatial noise and signal interference out, we threshold the probabilities $\bm{\rho}=tanh(\bm{P})$ at 0.5, discarding grids with $\rho_i<0.5$.
The final DoAs are then obtained by peak search on the filtered $\bm{P}$: the top $K$ peaks are taken when the number of sources is known; if fewer than $K$ peaks exist, the remaining directions are filled in descending order of energy.

\vspace{-0.15cm}
\section{Numerical Experiments}
\vspace{-0.2cm}
\subsection{Experiments Setup}
\vspace{-0.15cm}
The proposed model settings and experimental settings are listed in Table~\ref{tb:Experiments Setup}. The evaluation metric is RMSPE~\cite{merkofer2023music}. In the experiments, BeamformNet~\cite{deng2025beamformnet}, DA-MUSIC~\cite{merkofer2023music}, DeepSSE~\cite{xu2025deep}, SubspaceNet~\cite{shmuel2024subspacenet}, lowSNRNet~\cite{papageorgiou2021deep}, IQResNet~\cite{zheng2024deep}, and MUSIC~\cite{schmidt1986multiple} are adopted for comparison.

\begin{table}[!ht]
\scriptsize
\vspace{-0.5cm}
    \centering
    \caption{Experiments Setup}
    \label{tb:Experiments Setup}
    \begin{tabular}{|l|l|l|l|}
    \hline
        Parameter & Value & Parameter & Value \\ \hline
        $M$ & 16 & Training set size & $9\times10^4$ \\ 
        $c$ & 340 & Validation set size & $1\times10^4$ \\ 
        $f$ & 1000 & Test set size & $1\times10^4$ \\ 
        $\delta$ & 1° & Batch size & 32 \\ 
        $R$ & 180 & learning rate & $1\times10^{-4}$ \\ 
        $E$ & 256 & Optimizer & Adam \\ 
        Angle range & (-90°, 90°] & Epoch & 100 \\ 
        Array & ULA & Early stop epoch & 20 \\ \hline
    \end{tabular}
\vspace{-0.6cm}
\end{table}

\subsection{Results}
\vspace{-0.15cm}

We conducted experiments under various challenging scenarios:
Performance comparison across different numbers of sources at low SNR (SNR = 0 dB, T = 200), as shown in Fig.~\ref{fig:Experiments} (a). Performance comparison across different SNRs with a fixed number of sources (K = 4, T = 200), as shown in Fig.~\ref{fig:Experiments} (b). Performance comparison under limited snapshots with fixed SNR (K = 4, SNR = 20 dB), as shown in Fig.~\ref{fig:Experiments} (c). Performance comparison in coherent versus non-coherent environments (K = [2,4,6,8], T = 200, SNR = [-10,-5,0,5,10]), as shown in Fig.~\ref{fig:Experiments} (d). The above experiments demonstrate that our method achieves the optimal average performance across various SNRs and numbers of snapshots, particularly under fixed low SNR (SNR = 0 dB) conditions with varying numbers of sources. In the DoA estimation task, the proposed method attains the best performance in the single-snapshot scenario ($T = 1$), demonstrating its adaptability to limited snapshots. It also maintains satisfactory performance under extremely low SNR conditions (SNR $\leq-10$ dB), verifying its strong robustness.

\subsection{Ablation Study}
\vspace{-0.15cm}

We conducted three ablations: 1) path ablation (BX-C-Variant), comparing $\bm{BA}$ vs $\bm{BX}$ paths with the same covariance input; 2) input ablation (BA-X-Variant), comparing covariance vs raw $\bm{X}$ input under the same $\bm{BA}$ path; 3) joint ablation (raw $\bm{X}$ input and $\bm{BX}$ path), this variant is equivalent to BeamformNet, see Fig~\ref{fig:process}(b). Results (Fig.~\ref{fig:Experiments} (a-d)) show a decreasing average performance order among BeamFocusNet, BA-X-Variant, BX-C-Variant, and BeamformNet under low SNR and limited snapshots, confirming the advantages of explicit focusing of $\bm{BA}$ and covariance input, with explicit focusing contributing more.

We further ablate the embedding length (Fig.~\ref{fig:Experiments} (e)). Under the same embedding dimension, BeamFocusNet (Fig.~\ref{fig:process} (c)) outperforms BeamformNet (Fig.~\ref{fig:process} (b)), especially at low embedding dimensions and low parameter counts. This makes BeamFocusNet more suitable for capacity-constrained embedded devices compared to BeamformNet.

In addition, performance comparison in source number estimation at low SNR (for fair comparison, following~\cite{merkofer2023music} and~\cite{deng2025beamformnet}, a linear neural network is appended for classification), as shown in Fig.~\ref{fig:Experiments} (f) (SNR = 0 dB, T = 200). Experimental results show that our method is relatively weak at source number prediction. Ablation analysis indicates that applying filter $\bm{B}$ to the received signals $\bm{X}$ ($\bm{BX}$) is more suitable for source number estimation than applying it to $\bm{A}$ ($\bm{BA}$). We attribute this to $\bm{BX}$ providing a more robust active/inactive distinction: it implicitly suppresses noise, yielding lower and flatter output in inactive regions, whereas $\bm{BA}$ focuses on sharp, accurate angular peaks in active regions. We recommend using $\bm{BX}$ for source number estimation and $\bm{BA}$ for DoA estimation.

\vspace{-0.1cm}
\section{Conclusion}
\vspace{-0.25cm}
We propose the BeamFocusNet method for DoA estimation. It achieves the best average performance across various conditions and is more robust under low SNR and limited snapshots. During experiments, we find that the $\bm{BX}$ paradigm suits source number estimation, whereas the $\bm{BA}$ paradigm suits angle estimation. Future work will explore their fusion.

\vfill\pagebreak
{\fontsize{9}{9.2}\selectfont
\bibliographystyle{IEEEbib}
\bibliography{strings,refs}
}

\end{document}